\documentclass[aip, amsmath, amssymb, reprint,]{revtex4-1}
\usepackage{graphicx}
\usepackage{dcolumn}
\usepackage{bm}
\usepackage{physics}
\usepackage{amsfonts}
\usepackage{verbatim}
\usepackage{mathtools}
\usepackage{xcolor}
\usepackage{hyperref}
\hypersetup{
    colorlinks=true,
    linkcolor=blue,
    citecolor=blue,
    urlcolor=blue
}

\makeatletter
\def\@email#1#2{
    \endgroup
    \patchcmd{\titleblock@produce}
    {\frontmatter@RRAPformat}
    {\frontmatter@RRAPformat{\produce@RRAP{*#1\href{mailto:#2}{#2}}}\frontmatter@RRAPformat}
    {}{}
}
\makeatother

\begin{document}

\preprint{AIP/123-QED}

\title{Spatial and Temporal Control of Avalanches in Dense Active-Passive Mixtures}
\author{Caterina Landi}
\affiliation{Department of Structure of Matter, Thermal Physics, and Electronics, Universidad Complutense de Madrid, 2840 Madrid, Spain.}

\author{Giulia Janzen}
\affiliation{Department of Structure of Matter, Thermal Physics, and Electronics, Universidad Complutense de Madrid, 2840 Madrid, Spain.}

\author{Chantal Valeriani}
\affiliation{Department of Structure of Matter, Thermal Physics, and Electronics, Universidad Complutense de Madrid, 2840 Madrid, Spain.}

\author{Ezequiel E. Ferrero}
\affiliation{Instituto de Nanociencia y Nanotecnolog\'{\i}a, CNEA--CONICET, Centro At\'omico Bariloche, R8402AGP S. C. de Bariloche, R\'{\i}o Negro, Argentina.}
\email{ezequiel.ferrero@conicet.gov.ar}

\author{D. A. Matoz-Fernandez}
\affiliation{Department of Theoretical Physics, Universidad Complutense de Madrid, 2840 Madrid, Spain.}
\email{dmatoz@ucm.es}
\date{\today}

\begin{abstract}

Dense active--passive mixtures are often described within an effective-temperature picture, in which active dopants promote cage breaking and fluidize an arrested amorphous host.
However, at sufficiently large persistence, the same dopants drive stress accumulation and localized void formation. 
Here we investigate the avalanche dynamics across this progression, focusing on the intermediate regime between dynamic arrest and void formation, where the system remains structurally homogeneous on average but relaxes through intermittent collective rearrangements rather than as a simple activated liquid. Using overdamped simulations of dilute active dopants embedded in a dense amorphous solid, we show that the avalanche dynamics separates into distinct spatial and temporal controls. 
For a given system size, the active fraction sets the size of the largest stress-release events, with a cutoff that grows as the square root of the fraction of dopants, $S_c\sim \phi_a^{1/2}$, i.e., with the number of mean dopant spacings across the system, whereas persistence controls how avalanches unfold in time at fixed released stress. A decomposition of the stress into active and passive contributions, together with measurements of avalanche participation and morphology, shows that the active minority provides a localized and persistent driving channel while the surrounding amorphous packing carries the collective stress release. While the active minority controls the cutoff of the avalanche size statistics and the timing of the stress releases, the long-range elastic response and the building correlations in the passive host sets the size exponent, $\tau_S\simeq1.1$, that is shared with externally sheared and uniformly active amorphous solids.

\end{abstract}

\maketitle

%%%%%%%%%%%%%%%%%%%%%%%%%%%%%%%%%%%%%%%%%%%%%%%%%%%%%%%%%%%%
\section{Introduction}
%%%%%%%%%%%%%%%%%%%%%%%%%%%%%%%%%%%%%%%%%%%%%%%%%%%%%%%%%%%%

Dense active--passive mixtures provide a minimal model in which a localized non-equilibrium drive perturbs an otherwise arrested amorphous host, allowing the respective roles of activity and host mechanics to be disentangled~\cite{FreyPRL2016,ManningPRL2024}. They arise in driven colloids, granular media, bacterial suspensions, and living tissues, where constituents with heterogeneous dynamics interact in crowded, disordered environments~\cite{grosser21,coban21,dinelli2023non,grober2023unconventional}. In such mixtures the localized non-equilibrium forcing perturbs an otherwise glassy host, which raises the question of how activity drives structural relaxation and modifies mechanical rigidity near the glass transition~\cite{RevModPhys.88.045006,berthier2019glassy,janssen2019active}.

Motility heterogeneity has been studied from tracer dynamics in active or passive baths, where anomalous transport is reported~\cite{PhysRevLett.103.198103,C0SM00164C,PhysRevE.88.041002,mejia2011bias,benichou2013biased,Wittkowski_2017}, to dense active--passive mixtures that show cooperative motion and activity-induced rearrangements~\cite{vasilyev2017cooperative,Banerjee2022,Lozano2019,PhysRevLett.122.148101,D1SM01092A}. In dense regimes these mixtures display the crystallization of hard-sphere glasses~\cite{Ni2014_Crystallizing}, the segregation of anisotropic particles~\cite{C2SM06960A}, and activity-induced phase separation and collective flow~\cite{PhysRevLett.112.198301,PhysRevLett.114.018301,C4SM01562B}. More recent work addresses the dilute limit, in which a small fraction of active dopants alters the mechanical response and relaxation pathways of an otherwise passive amorphous host~\cite{Paoluzzi2024,omar2018swimming}.

Activity in dense amorphous media is often treated as an effective temperature that facilitates relaxation and drives glass--fluid transitions~\cite{mandal2016active,MatozPRL2017,Shee2025dopants, Tahaei2026}. 
This mapping is not universal. It breaks down when the active forces stay correlated over times comparable to or longer than the mechanical relaxation time of the host, where the injected stress becomes a structured, long-lived perturbation rather than isotropic noise~\cite{levis2015single,PetrellieffectiveT2020,mandal2020multiple,JanzenAging2022,WieseKroyLevis2023}. Driven intruders make this mechanical origin explicit, generating persistent structural defects in the surrounding medium rather than amplifying thermal fluctuations~\cite{Reichhardt2004PhysRevLett}. Dense active matter thus connects to athermally driven disordered solids, with persistent forces perturbing the energy landscape rather than acting as noise~\cite{Morse2021,PhysRevFluids.2.043103,PhysRevE.92.032118,Yanagishima2017}. Once the persistence time exceeds the elastic relaxation time~\cite{Matoz2020}, injected stresses do not fully relax between reorientations and accumulate and spread over mesoscopic distances.
Whether persistence, rather than an effective temperature, is the key ingredient for stress accumulation and scale-free relaxation can be tested directly by comparing active and thermal dopants within the same host.

In a dilute active--passive mixture, at moderate persistence the dopants fluidize the arrested host, whereas sufficiently persistent forcing produces stress accumulation, spatial localization, and void formation~\cite{Janzen2026}. 
The apparently homogeneous fluidized state between arrest and void formation is not necessarily a conventional activated liquid. 
It may retain the intermittent mechanical response of an amorphous solid, relaxing through localized yielding and correlated stress-release 
cascades, as happens even in supercooled liquids~\cite{ChattorajPRL2013}.

Avalanche dynamics are well established in externally driven amorphous solids~\cite{Nicolas2018}. 
A local plastic instability generates an anisotropic, long-ranged elastic disturbance that destabilizes neighboring regions and produces a collective cascade~\cite{FalkLanger1998,MaloneyLemaitre2006,MaloneyRobbins2008,LinPNAS2014, Sanz2014,Liu2016,Nicolas2018,Yanagishima2023}. 
Closely related avalanche statistics have been reported under active and spatially heterogeneous driving~\cite{Morse2021,MandalSollich2021,Villarroel2024,Roland2026}. The similarity of the exponents across protocols, however, hides the distinct roles of the drive and of the matrix. Different protocols determine how local yielding is initiated, but the subsequent propagation may be governed mainly by the mechanical response of the amorphous packing. Active--passive mixtures separate these two roles directly: an active minority supplies a controlled, spatially distributed source of persistent forcing, which at low temperatures is the dominant and only energy injection source in the system, while a passive majority forms the mechanically arrested host through which stress is transmitted and spatial correlations built. We can therefore ask whether activity/persistence controls how local yielding events are generated while the host controls the spatial extent and scale-free statistics of the resulting stress release.

Here, we numerically investigate a two-dimensional dense active--passive mixture across the three dynamical regimes identified in ~\textcite{Janzen2026}: dynamic arrest, homogeneous fluidization, and void formation. We characterize the global stress signal, the size-conditioned distributions, the durations at fixed size, the active and passive stress contributions, and the spatial extent of particle rearrangements as functions of the active fraction and persistence length. These two control parameters act on different aspects of the dynamics. 
The active fraction sets the spatial scale of the largest stress releases through the mean spacing between dopants, whereas persistence controls their initiation and timing. The propagation mechanism itself is governed by the elastic mechanics of the amorphous host.
We further show that replacing active dopants with thermal ones yields purely exponential statistics throughout, establishing that persistent forcing coming from the active particles, rather than an effective temperature replacement of them, is responsible for the scale-free dynamics.

Unlike protocol-comparison studies, which impose a \emph{global} drive on a uniformly active or sheared system~\cite{Morse2021,Villarroel2024,Roland2026}, we realize the drive/host separation with two physically distinct populations, so that the forcing (the active minority) and the mechanically responding medium (the passive majority) are resolved within a single run. 
This geometry lets us define an avalanche-size cutoff that at at fix system size scales with the fraction of dopants, $S_c\sim \phi_a^{1/2}$, a control that a uniformly driven system cannot access, and we verify its length-controlled origin by an independent finite-size scan (Supplementary Material~\cite{SI}). Relative to the companion phase-diagram study of Ref.~\cite{Janzen2026}, which maps the fluidization and void-formation boundaries, we resolve the avalanche dynamics \emph{across} these regimes and show that their scaling exponents are host-controlled and essentially very similar from non-active systems, whereas the persistent activity controls their size cutoff and timing.

%%%%%%%%%%%%%%%%%%%%%%%%%%%%%%%%%%%%%%%%%%%%%%%%%%%%%%%%%%%%
\section{Model and Simulation Details}
\label{sec:Model}
%%%%%%%%%%%%%%%%%%%%%%%%%%%%%%%%%%%%%%%%%%%%%%%%%%%%%%%%%%%%

We simulate a two-dimensional dense amorphous packing of $N=10^4$ soft disks at area fraction $\phi=0.9$, well inside the mechanically crowded regime. Crystallization is suppressed by assigning each particle a radius $a_i$ drawn from a uniform distribution with mean $a$ and half-width $0.2\,a$.
Particles interact through a finite-range purely repulsive harmonic contact force: $\mathbf F_{ij}=k(a_i+a_j-r_{ij})\hat{\mathbf r}_{ij}$ for $r_{ij}<a_i+a_j$ and $\mathbf F_{ij}=\mathbf 0$ otherwise, where 
$r_{ij} = |\mathbf{r}_i - \mathbf{r}_j|$ and 
$\hat{\mathbf{r}}_{ij} = (\mathbf{r}_i-\mathbf{r}_j)/r_{ij}$.
The initial condition is a dense disordered configuration of passive particles, from which we randomly select $N_a$ particles and assign them self-propulsion. We define the active fraction as $\phi_a = N_a/N$.

Each particle $i$ is described by its position vector $\mathbf r_i$ and orientation vector $\mathbf n_i=(\cos\theta_i$, $\sin\theta_i)$, with $\theta_i$ measured with respect to the laboratory $x$ axis. Positions and orientations evolve according to the overdamped Langevin equations:
  \begin{align}
  \dot{\mathbf r}_i &= \mu\sum_{j\neq i}\mathbf F_{ij} + v_0^{(i)}\,\mathbf n_i
                     + \sqrt{2D_t}\,\boldsymbol\xi_i(t), \label{eq:eom_r}\\
  \dot\theta_i &= \sqrt{2D_r^{(i)}}\,\eta_i(t), \label{eq:eom_theta}
  \end{align}
where $\mathbf F_{ij}$ is the contact force defined above, $\mu$ is the translational mobility, $v_0^{(i)}$ is the self-propulsion speed, $D_t$ and $D_r^{(i)}$ are the translational and rotational diffusion coefficients, and $\boldsymbol\xi_i(t)$ and $\eta_i(t)$ are Gaussian white noises with zero mean and unit variance, uncorrelated between different particles and Cartesian components. Active particles self-propel with speed $v_0^{(i)}=v_0$ and undergo rotational diffusion with coefficient $D_r^{(i)}=D_r$. Since no torques or alignment interactions are present, their orientations decorrelate through rotational diffusion alone, with persistence time $\tau_r=D_r^{-1}$ and persistence length $l_p=v_0\tau_r$. Passive particles have $v_0^{(i)}=0$ and $D_r^{(i)}=0$, so they move only under interparticle forces and weak translational noise, and their orientations remain constant.

All quantities are expressed in simulation units, in which the mean particle radius $a$, the mobility $\mu$ and the self-propulsion speed $v_0$ are set to one, so that lengths are measured in units of $a$ and times in units of $\tau_0=a/v_0$. The contact stiffness is set to $k=10$, so that the elastic relaxation time is $(\mu k)^{-1}=0.1\,\tau_0$. We set $D_t=10^{-7}$ to regularize the overdamped dynamics without perturbing the structure of the passive packing. 
We vary the persistence length $l_p\in[2\times10^{-1},\,2\times10^{3}]$ by tuning $D_r\in[5\times10^{-4},\,5\times10^{0}]$. We also vary the active fraction $\phi_a\in[10^{-4},\,10^{-1}]$, with the lowest value corresponding to a single active particle ($N_a=1$). Unless otherwise stated, results are presented along two complementary cuts chosen to probe the system across its different dynamical regimes (see Fig.~\ref{fig:phase_diagram} below): varying $l_p$ at fixed $\phi_a=10^{-1}$, or varying $\phi_a$ at fixed $l_p=10$.

The equations of motion are integrated with a time step $\mathrm{d}t=5\times10^{-3}\,\tau_0$ using the GPU-accelerated \textsc{SAMoS} simulation package~\cite{SAMoS2024}. After letting the system evolve for at least $10^5\,\tau_0$, particle positions and the stress defined in Eq.~\eqref{eq:total_stress_signal} are recorded over a time window of $10^5\,\tau_0$. Positions are saved every $\tau_0$, whereas the stress is sampled every $0.05\,\tau_0$. The finer temporal resolution is required to resolve individual stress drops, while storing particle configurations at the same rate would be prohibitively expensive.

\begin{figure*}
    \centering
    \includegraphics[width=0.97\linewidth]{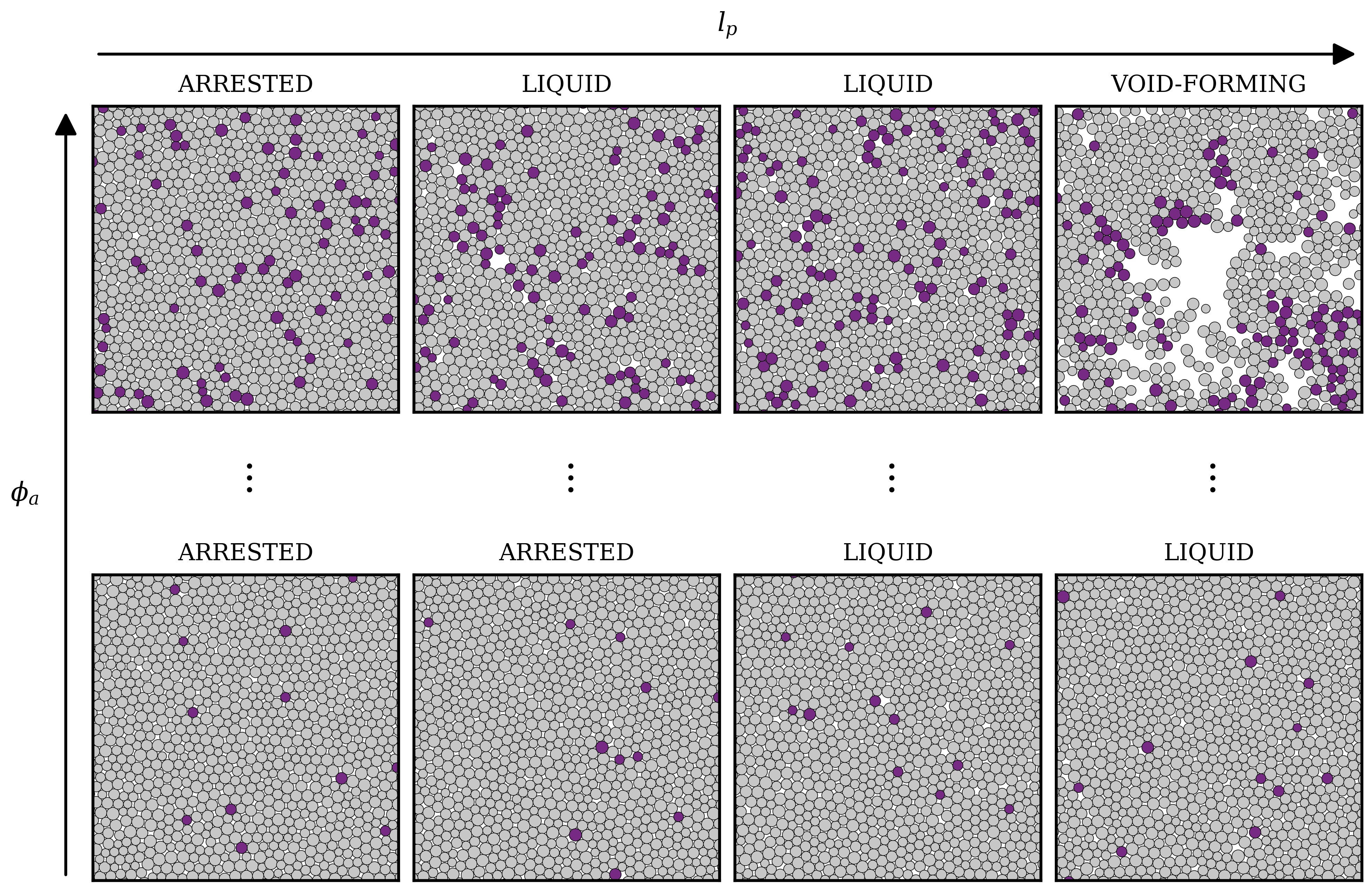}
    \caption{Snapshots of the final configurations for active fractions $\phi_a = 10^{-1}$ (top row) and $10^{-2}$ (bottom row) and persistence lengths $l_p = 2\times10^{-1},\ 10,\ 2\times10^{2},$ and $2\times10^{3}$ (from left to right).
    The top row illustrates the three dynamical regimes: the arrested state ($l_p = 2\times10^{-1}$), the homogeneous fluid ($l_p = 10$ and $2\times10^{2}$, the latter close to the onset of the void-forming regime~\cite{Janzen2026}), and the void-forming regime ($l_p = 2\times10^{3}$). The bottom row shows that the arrest--fluidization transition is shifted to larger $l_p$ at a lower $\phi_a$: the system is arrested at $l_p = 2\times10^{-1}$ and $10$, and fluidizes at $l_p = 2\times10^{2}$ and $2\times10^{3}$, the former near the arrest--fluidization transition line~\cite{Janzen2026}. No void formation is observed in the investigated range. The dots indicate intermediate values of $\phi_a$. Lower active fractions, down to $\phi_a = 10^{-4}$, are included in the rest of the analysis but not shown here for clarity. Active particles are shown in purple and passive particles in gray, all drawn with their actual radii. Each panel shows a zoomed-in region of the simulation box. The corresponding videos are provided in the Supplementary Material~\cite{SI}.}
    \label{fig:phase_diagram}
\end{figure*}

%%%%%%%%%%%%%%%%%%%%%%%%%%%%%%%%%%%%%%%%%%%%%%%%%%%%%%%%%%%%
\section{Observables}
\label{sec:observables}
%%%%%%%%%%%%%%%%%%%%%%%%%%%%%%%%%%%%%%%%%%%%%%%%%%%%%%%%%%%%

%%%%%%%%%%%%%%%%%%%%%%%%%%%%%%%%%%%%%%%%%%%%%%%%%%%%%%%%%%%%
\subsection{Stress Drops}
\label{sec:stress_signal}
%%%%%%%%%%%%%%%%%%%%%%%%%%%%%%%%%%%%%%%%%%%%%%%%%%%%%%%%%%%%

In dense amorphous systems, deformation occurs through localized structural changes known as plastic rearrangements, in which particles change their local neighbor configurations \cite{FalkLanger1998}. When the system is loaded, it deforms elastically, accumulating stress until a local rearrangement takes place, triggering a cascade of consecutive and correlated plastic events among neighboring particles. Such a collective sequence of events is termed an ``avalanche'' and macroscopically manifests as a sudden drop in the shear stress \cite{Nicolas2018}. 

The stress tensor is computed using the Irving--Kirkwood formula \cite{irving1950}: 
\begin{equation}
\sigma_{\alpha\beta} = \frac{1}{2L^2} \sum_{i\neq j} r_{ij}^{\alpha}F_{ij}^{\beta},
\label{eq:total_stress_signal}
\end{equation}
where Greek indices $\alpha$ and $\beta$ label Cartesian components, Latin indices $i$ and $j$ label particles, and $L$ is the edge length of the simulation box. Eq.~\eqref{eq:total_stress_signal} is the interaction (virial) stress only: it contains only the pairwise contact forces and not the self-propulsion forces. This is the relevant quantity because plastic avalanches correspond to sudden relaxations of the elastic energy stored in the contact network, while the self-propulsion acts as the external drive.

In the following, we focus on the shear component $\sigma(t)\equiv\sigma_{xy}(t)$. To isolate individual avalanches, we monitor the negative excursions of the shear stress rate, i.e., $\dot{\sigma} = \mathrm{d}\sigma/\mathrm{d}t < 0$, following Refs.~\cite{Roland2026, Liu2016}. 
Specifically, a single avalanche is defined as a contiguous interval of negative stress slope, bounded by an onset time $t_i$ (where $\dot{\sigma}$ becomes negative) and an end time $t_f$ (where $\dot{\sigma}$ becomes positive again). To ensure the robustness of this definition, a sensitivity analysis based on a detection threshold is reported in the Supplementary Material \cite{SI}.

For each detected avalanche, its duration is defined as $T = t_f - t_i$, and its size $S$ as the extensive total stress released during the drop:
\begin{equation}
S = N \Delta \sigma, 
\end{equation}
where $\Delta \sigma = \sigma(t_i) - \sigma(t_f)$ and $N$ represents the number of particles in the system. From the ensemble of avalanches collected, we construct the probability distributions of durations $p(T)$ and sizes $p(S)$.

To resolve the contribution of each species to the avalanches, we split the stress of Eq.~\eqref{eq:total_stress_signal} into the terms associated with active ($\nu = A$) and passive ($\nu = P$) particles:
\begin{equation}\label{eq:sigma_nu}
%\sigma^\nu_{\alpha\beta} = \frac{1}{2L^2}\sum_{j\in\nu}\sum_{i\neq j}
\sigma^\nu_{\alpha\beta} = \frac{1}{2L^2}\sum_{j\in\nu ~\&~i\neq j}
r^\alpha_{ij}F^\beta_{ij},
\end{equation}
where the sum over $j$ runs over the particles of species $\nu$ only. Pairs of particles of the same species contribute fully to $\sigma^\nu_{\alpha\beta}$, whereas each mixed pair contributes half to each species, so that $\sigma_{\alpha\beta} = \sigma^A_{\alpha\beta} + \sigma^P_{\alpha\beta}$ exactly. From here on, $\sigma^\nu(t) \equiv \sigma^\nu_{xy}(t)$, in the same notation as the total stress.

We measure the contribution of both species to each avalanche over the \emph{same} time window identified from the total stress signal. Detecting avalanches independently in $\sigma^A$ and $\sigma^P$ would instead yield two event lists that cannot be matched to each other. With this common window, the stress drop of each species is:
\begin{equation}
S^\nu = N\Delta\sigma^\nu,
\qquad
\Delta\sigma^\nu = \sigma^\nu(t_i) - \sigma^\nu(t_f),
\end{equation}
so that $S = S^A + S^P$ for every avalanche. While $S > 0$ by construction (since $\Delta\sigma>0$ by definition of a stress drop), the partial sizes $S^\nu$ can take either sign: $S^\nu > 0$ means that species $\nu$ releases stress during the avalanche, while $S^\nu < 0$ means that it is loaded while the system as a whole relaxes. Restricting to the avalanches with $S^\nu > 0$, we construct the partial size distributions $p(S^A)$ and $p(S^P)$, each normalized to unity over its own set of events.

%%%%%%%%%%%%%%%%%%%%%%%%%%%%%%%%%%%%%%%%%%%%%%%%%%%%%%%%%%%%
\subsection{Participation Ratio and Mobile Clusters}
%%%%%%%%%%%%%%%%%%%%%%%%%%%%%%%%%%%%%%%%%%%%%%%%%%%%%%%%%%%%

To quantify the spatial extent of rearrangements, we compute the participation ratio~\cite{Roland2026}:
\begin{equation}
P = \frac{ \left( \sum_i d_i^2 \right)^2 }{ N \sum_i d_i^4 },
\end{equation}
where $d_i$ represents the displacement magnitude of particle $i$ during an avalanche. Specifically, having identified the initial and final times of a given avalanche $(t_i, t_f)$, the displacement vector components are calculated directly from the particle positions as:
\begin{align}
\Delta x_i &= x_i(t_f) - x_i(t_i), \\
\Delta y_i &= y_i(t_f) - y_i(t_i),
\end{align}
and the displacement magnitude is $d_i = \sqrt{\Delta x_i^2 + \Delta y_i^2}$. 

The participation ratio $P$ ranges between $1/N$, corresponding to a highly localized rearrangement where a single particle moves while the rest of the system remains still, and $1$, representing an idealized homogeneous scenario where all $N$ particles undergo identical displacements. 
This formulation provides a robust measure of the spatial extent of the rearrangements without requiring any arbitrary threshold on the particle displacements. 

Using this property, we systematically identify the set of `mobile particles' by selecting the top $\lfloor NP \rfloor$ particles exhibiting the largest displacements $d_i$ within the considered avalanche. 

A `mobile cluster' is then defined as a spatially connected component of mobile particles, where two particles are considered connected if their distance is smaller than the average particle diameter $2a$. These clusters represent the regions where the structural rearrangement takes place, and their analysis allows us to quantify the geometry of these rearranging regions through their mass--radius scaling.

%%%%%%%%%%%%%%%%%%%%%%%%%%%%%%%%%%%%%%%%%%%%%%%%%%%%%%%%%%%%
\section{Results}
\label{sec:results}
%%%%%%%%%%%%%%%%%%%%%%%%%%%%%%%%%%%%%%%%%%%%%%%%%%%%%%%%%%%%

%When a dilute set of persistent dopants relaxes an arrested amorphous host, which features of the collective response are set by the \emph{drive} and which are set by the \emph{host}? In driven amorphous solids the forcing protocol and the mechanical response of the packing play separable roles. Protocols that interpolate between uncorrelated local forcing and system-spanning shear yield critical responses that map onto one another up to a rescaling set by the correlation length of the applied field~\cite{Morse2021}. Active and shear driving produce avalanche exponents that agree within uncertainty despite different steady-state rheology~\cite{Villarroel2024,Roland2026}. Independently of the drive, the mechanical response is set by the host alone. The elementary plastic event is a localized non-affine rearrangement embedded in an elastic medium~\cite{FalkLanger1998}. Its Eshelby-like stress field redistributes stress anisotropically over long ranges, organizing subsequent yielding into correlated, often line-like, cascades that are largely insensitive to microscopic interaction details~\cite{MaloneyLemaitre2006,MaloneyRobbins2008,Nicolas2018}. Our active--passive mixture realizes this separation of roles at the particle level. The active population fixes \emph{where and when} local yielding nucleates and temporally organizes, while the amorphous host governs \emph{how} the resulting stress release propagates.

We investigate the separate roles of the \emph{drive} and the \emph{host} in shaping the collective response of an arrested amorphous solid doped with a dilute set of persistent particles. In driven amorphous solids, the forcing protocol and the mechanical response of the packing play separable roles: the statistical features of the response are set by the host alone, independently of the drive. Protocols that interpolate between uncorrelated local forcing and system-spanning shear yield critical responses that map onto one another up to a rescaling set by the correlation length of the applied field~\cite{Morse2021}, and active and shear driving produce avalanche exponents that agree within uncertainty despite different steady-state rheology~\cite{Villarroel2024,Roland2026}. This insensitivity to the drive reflects the nature of the elementary plastic event: a localized non-affine rearrangement embedded in an elastic medium~\cite{FalkLanger1998}, whose Eshelby-like stress field redistributes stress anisotropically over long ranges, organizing subsequent yielding into correlated, often line-like, avalanches~\cite{MaloneyLemaitre2006,MaloneyRobbins2008,LinPNAS2014, Liu2016,Nicolas2018}. Yet precisely what role the drive plays in organizing this plastic response remains an open question. Our active--passive mixture makes this separation explicit at the particle level: the active population fixes \emph{where and when} local yielding is nucleated, while the amorphous host governs \emph{how} the resulting stress release propagates through such cascades.

Before analyzing the avalanche statistics, we place the control parameters used in the following within the phase diagram established for this model. Previous work on the same system showed that the persistence length $l_p$ and the active fraction $\phi_a$ give rise to three distinct dynamical regimes in the $(l_p,\phi_a)$ plane: a dynamically arrested state, a spatially homogeneous fluid, and a void-forming regime at sufficiently strong persistent forcing~\cite{Janzen2026}. Representative configurations are shown in Fig.~\ref{fig:phase_diagram} (see also the corresponding videos in the Supplementary Material~\cite{SI}), and we use this map to sample the three regimes along two complementary cuts.

At fixed $\phi_a = 10^{-1}$ (top row of Fig.~\ref{fig:phase_diagram}), increasing $l_p$ from $2\times10^{-1}$ to $2\times10^{3}$ drives the system successively through the arrested state, the homogeneous fluid, and finally the void-forming regime; for $l_p = 10$ the system is already a homogeneous fluid, and $l_p = 2\times10^{2}$ lies close to the onset of void formation~\cite{Janzen2026}. Complementarily, at fixed $l_p = 10$ (second column of Fig.~\ref{fig:phase_diagram}) we vary $\phi_a$. Along this cut the system crosses the arrest--fluidization boundary: it is a homogeneous fluid at $\phi_a = 10^{-1}$ and already arrested at $\phi_a = 10^{-2}$, so that lower fractions, down to $\phi_a = 10^{-4}$, only drive it deeper into the glassy state. Here ``arrested'' refers to the absence of structural relaxation within the observation window; plastic rearrangements and stress drops still occur in this state, so the avalanche statistics below also sample the arrested side of the boundary.

We characterize the contribution of each population across the three dynamical regimes. In the dilute regime, the active dopants fluidize the arrested host. In the homogeneous fluid regime, the active dopants remain more mobile than the passive medium. At strong persistence, the mobilities of the two populations become comparable, and rearrangements localize at the boundaries of self-generated voids.

%%%%%%%%%%%%%%%%%%%%%%%%%%%%%%%%%%%%%%%%%%%%%%%%%%%%%%%%%%%%
\subsection{Stress Signal}

\begin{figure}[t!]
    \centering
    \includegraphics[width=\columnwidth]{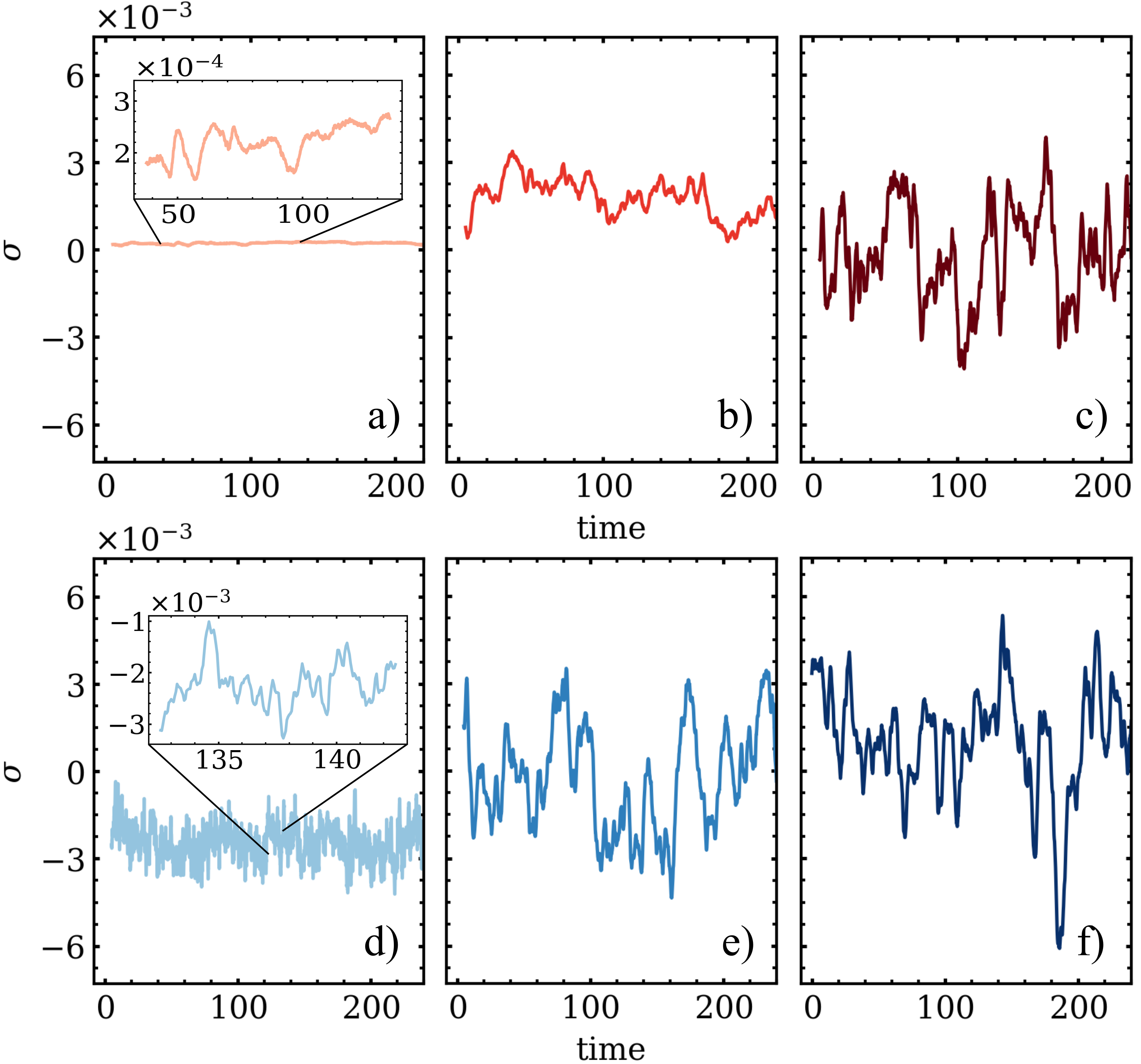}
    \caption{Stress time series varying the fraction of active dopants at fixed persistence length, $l_p = 10$. (a) $\phi_a = 10^{-4}$, (b) $\phi_a = 10^{-2}$, and (c) $\phi_a = 10^{-1}$; or varying the persistence length at fixed fraction of active dopants, $\phi_a = 10^{-1}$: (d) $l_p = 2 \times 10^{-1}$, (e) $l_p = 2 \times 10^1$, and (f) $l_p = 2 \times 10^3$. Insets in (a) and (d) show zoomed views. Color coding is consistent with Figs.~\ref{fig:avalanche_size_distributions} and~\ref{fig:avalanche_duration_distributions}.}
    \label{fig:stress_timeseries}
\end{figure}

Figure~\ref{fig:stress_timeseries} shows representative steady-state shear stress time series at different active fractions $\phi_a$ and persistence lengths $l_p$.

All series display intermittent dynamics, in which stress is repeatedly built up and released in discrete events.
Such bursts are the signature of plastic avalanches in an amorphous solid, in which localized rearrangements redistribute stress and trigger correlated yielding in neighboring regions~\cite{Nicolas2018}. 
Unlike in sheared amorphous solids~\cite{Nicolas2018}, where slow loading is punctuated by sudden stress drops, here we find no significant asymmetry between stress rises and drops (see Supplementary Material~\cite{SI}).
This intermittency persists even though the drive is a dilute persistent minority rather than an imposed global strain, consistent with observations in fully active glasses where persistent forcing similarly drives intermittent Eshelby-like plastic events~\cite{MandalSollich2021, Roland2026}.

Focusing on $\phi_a$ at fixed $l_p$ (top panels), we observe a clear effect on the drop amplitude. For low doping, $\phi_a = 10^{-4}$ [Fig.~\ref{fig:stress_timeseries}(a)], the signal appears nearly flat, although the inset reveals that the characteristic drop-like structure is already present. As $\phi_a$ increases to $10^{-1}$ [Fig.~\ref{fig:stress_timeseries}(c)], the amplitude grows. The larger drops follow from the greater mechanical energy injected by a denser active population, in which more triggering sites accumulate and release more stress.

Varying $l_p$ at fixed $\phi_a$ (bottom panels) affects the amplitude similarly, but here the temporal structure also changes visibly. At low $l_p$ [Fig.~\ref{fig:stress_timeseries}(d)], the dynamics consist of a rapid succession of tiny drops (inset), whereas at $l_p = 2\times10^3$ [Fig.~\ref{fig:stress_timeseries}(f)] the drops become larger and more widely spaced, indicating longer loading intervals. Both parameters thus increase the drop amplitude and lengthen the loading periods, through microscopic routes that the avalanche statistics below separate. The total number of drops and the distribution of waiting times are reported in the Supplementary Material~\cite{SI}.

%%%%%%%%%%%%%%%%%%%%%%%%%%%%%%%%%%%%%%%%%%%%%%%%%%%%%%%%%%%%
\subsection{Avalanche Size Distributions}
%%%%%%%%%%%%%%%%%%%%%%%%%%%%%%%%%%%%%%%%%%%%%%%%%%%%%%%%%%%%

\begin{figure}[t!]
    \centering
    \includegraphics[width=\columnwidth]{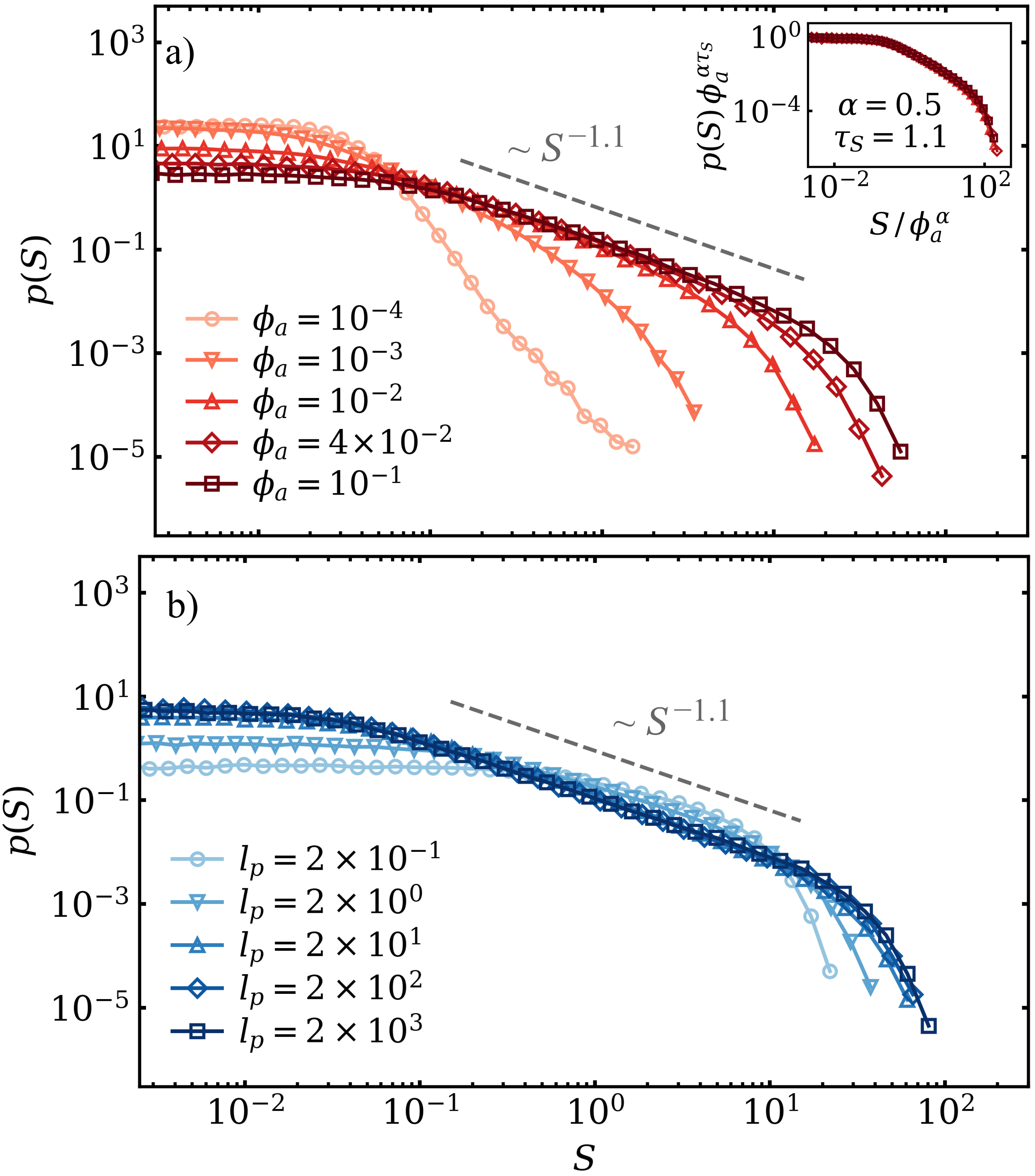}
    \caption{Avalanche size distributions $p(S)$ upon varying $\phi_a$ at fixed $l_p = 10$ (top panel), and upon varying $l_p$ at fixed $\phi_a = 10^{-1}$ (bottom panel). 
    Increasing either $\phi_a$ or $l_p$ drives the distributions from an exponential decay toward a power-law $p(S) \propto S^{-\tau_S}$ with $\tau_S \simeq 1.1$ (dashed gray lines). Inset of top panel: scaling collapse of the power-law distributions ($\phi_a = 10^{-2}, 4 \times 10^{-2},$ and $ 10^{-1}$) obtained by plotting $p(S) \, \phi_a^{\alpha\tau_S}$ versus $S / \phi_a N^{\alpha}$ with $\alpha=0.5$. Distributions are computed and normalized over avalanches with $S > 5\times10^{-4}$.}
    \label{fig:avalanche_size_distributions}
\end{figure}

The size distributions $p(S)$ (Fig.~\ref{fig:avalanche_size_distributions}) change shape with 
the activity control parameters. 
Increasing either $\phi_a$ at fixed $l_p$ (top panel) or $l_p$ at fixed $\phi_a$ (bottom panel) 
drives a transition from an exponential decay toward a power-law distribution, eventually 
terminated by a faster-decaying cutoff function
\begin{equation}
 p(S)\propto S^{-\tau_S} \mathcal{F}(S/S_c),
\end{equation}
with $\tau_S\simeq1.1$ and $S_c \equiv S_c(L,\phi_a,l_p)\sim L^{d_f}\phi_a^{\alpha}l_p^h$ the cutoff size. Here $L=\sqrt{N}$, the linear system size.
As a matter of fact, the distribution cutoff can have \textit{a priori} multiple dependencies. 
As shown in Fig.~\ref{fig:avalanche_size_distributions}(a), decreasing $\phi_a$ from $10^{-1}$ to $10^{-2}$ progressively narrows the power-law region and shifts the upper cutoff toward smaller $S$. 
At $\phi_a=10^{-4}$ the power-law is lost entirely and the distribution reverts to a 
pure exponential. 
This indicates that for such small active populations, correlations between plastic rearrangements are not easily established and the response stays confined to localized events.
In other words, a few scattered active particles act no differently than a weak thermal noise.
Moreover, replacing active dopants with thermal ones, i.e., considering a hot-cold mixture, yields purely exponential $p(S)$ (and also $p(T)$) for all temperatures tested (see Supplementary Material~\cite{SI}).
In other words, the building up of long-range correlations between plastic events and the power-laws emergence appears with enough persistent activity, coherent local strain accumulation, and not simple agitation.

As shown in the inset of Fig.~\ref{fig:avalanche_size_distributions}(a), the power-law curves ($\phi_a\ge10^{-2}$) collapse onto a single master curve upon rescaling $S/(\phi_a)^\alpha$ and $p(S)(\phi_a)^{\alpha\tau_S}$ with $\alpha=0.5$, so that the maximum avalanche size scales as $S_c\sim(\phi_a)^{1/2}$. That is, for a given system size, the active fraction controls the maximum event size through the cutoff.
In contrast, Fig.~\ref{fig:avalanche_size_distributions}(b) shows that for $l_p\ge2\times10^1$ the distributions rapidly reach the same power-law and that, for these values, the cutoff is essentially independent of $l_p$, all curves truncating near a common value, so the cutoff scaling exponent with $l_p$ being $h\simeq 0$. 
The two control parameters therefore act on different aspects of the dynamics. A sufficiently long directional push (large $l_p$) is required to break particles out of their cages and initiate the scale-free regime, but persistence alone cannot enlarge the spatial reach of a cascade. Once a cascade is triggered, the size of the largest events is controlled by the density of active sources rather than by $l_p$, with a cutoff set by the number of mean dopant spacings across the system, $L/\ell_a$, where $\ell_a\sim N_a^{-1/2}L$ (see below).
Finally, when varying the system size at fixed $\phi_a$ and $l_p$, we observe the typical finite-size scaling of $P(S)$ with a cutoff growing as $S_c \sim L^{d_f}$, as discussed below.

The measured exponent $\tau_S\simeq1.1$ lies within the range reported for driven disordered solids, being close to the athermal-quasistatic values of Morse \emph{et al.}~\cite{Morse2021} ($\tau_S\simeq1.0$) and Villarroel and D\"uring~\cite{Villarroel2024} ($\tau_S\simeq1.14$), somewhat above the value for non-active sheared systems~\cite{Ruscher2021} ($\tau_S\simeq0.98$), and consistent with the systematic increase from shear-dominated ($\tau_S\simeq1.0$) to activity-dominated ($\tau_S\simeq1.2$) driving documented by Wiese \emph{et al.}~\cite{Roland2026}. The proximity of our exponent to values reported under both shear and active driving should not be read as identifying an intermediate universality class. Avalanche exponents in particle-based models display appreciable protocol dependence, varying with the driving mechanism~\cite{Roland2026,Villarroel2021}, the driving rate~\cite{Liu2016}, and the damping and dynamical rule~\cite{Salerno2012,SalernoRobbins2013, Karimi2013}, whereas the local organization of plastic rearrangements and the signatures of long-range elastic stress redistribution are more robust~\cite{MaloneyLemaitre2006,MaloneyRobbins2008,FerreroJagla2019}. 
The weak variation of the size exponent across these protocols, including ours, is what one expects if the exponent is set by the elastic response of the host.

\subsubsection*{Discussion on the avalanche size distribution cutoff}

In the avalanche statistics of amorphous solids, in contrast with depinning-like systems where
a positively-defined interaction induces compact avalanches and so $d_f=d$, the Eshelby-like 
stress propagator induce correlation growing mostly along its strong interaction lines and yields \begin{equation}
S_c\sim L^{d_f} 
\end{equation} 
with avalanche fractal dimension $d_f\simeq 1$~\cite{LinPNAS2014,Nicolas2018}.

We confirm this scaling in our system independently through a finite-size analysis. 
Varying the total system size $N$ at fixed active fraction collapses the 
size distributions under the finite-size scaling ansatz 
$p(S)=S^{-\tau_S}\,\mathcal{F}\!\left(S/N^{d_f/2}\right)$ with an avalanche 
fractal dimension $d_f\simeq1$ (Supplementary Material~\cite{SI}), i.e., $S_c\sim N^{1/2}$.

Now, with this, line-like geometry of the avalanches at hand, the square-root dependence of 
the cutoff controlled by $\phi_a$ can be understood from the geometry. 
Let the mean spacing between active dopants be
\begin{equation}
    \ell_a \sim \rho_a^{-1/2} \sim \frac{L}{\sqrt{N_a}},
\end{equation}
where $\rho_a = N_a / L^2$ is the active dopant number density, the cutoff can be rewritten as
\begin{equation}
    S_c \sim N_a^{1/2} \sim \frac{L}{\ell_a},
\end{equation}
so the largest stress-release events scale with the number of active
\emph{spacings} across the system rather than with the total active population.
This counting has a simple mechanical interpretation. 
Each dopant stores a local prestress of order the active force, $f_a=v_0/\mu$, acting over a particle size $a$. 
When it relaxes, this produces a bond-level stress contribution distributed over the box. 
In the extensive units used for $S$, the corresponding stress release is
\begin{equation}
    S_{1} \sim N\,\frac{f_a a}{L^2} \sim \rho\, f_a a ,
\end{equation}
which is of order unity in our units and independent of $N_a$ and $l_p$. The finite-size analysis shows that the largest avalanches are line-like, so a system-spanning event is a backbone of length of order $L$. Such a backbone crosses $L/\ell_a=N_a^{1/2}$ dopants. The passive host propagates the release through its long-range elastic response, but it can only relay stress that has been stored, and the stored stress sits at the dopants. The largest release is therefore the number of dopants on the backbone times the release per dopant, $S_c\sim (L/\ell_a)\,S_1\sim N_a^{1/2} \sim L \phi_a^{1/2}$, with no adjustable parameter. 
A compact two-dimensional cascade would instead collect all dopants in a correlated area and give $S_c\sim (L/\ell_a)^2\sim N_a$, which the data rule out. 
The same argument explains why $S_c$ does not move with $l_p$ in Fig.~\ref{fig:avalanche_size_distributions}(b), that is, persistence changes neither the stress stored per dopant nor the number of dopants a spanning line crosses. 
It also anticipates the active--passive decomposition reported below, since each dopant releases only its own neighborhood whereas the host releases the elastic field that every local release drags along, so the active partial distribution must be cut off before the passive one.

The observed length scaling points to an effectively lower-dimensional, line-like propagation of stress release through the host, consistent with the slip-line and shear-transformation-chain organization established for sheared amorphous solids~\cite{MaloneyLemaitre2006,MaloneyRobbins2008}, with the active dopants acting only as spatially distributed cage-breaking seeds. That a dilute perturbation suffices to trigger system-scale release is expected for a marginally stable host, whose response to nominally small perturbations is singular and scale-free~\cite{Shang2020,Yanagishima2021}. The active particle does not need to supply the avalanche energy, only destabilize a host already close to yielding. The scaling $S_c\sim N_a^{1/2}$ is not by itself proof of strictly one-dimensional avalanches. It shows that the stress-drop cutoff is set by a length rather than by a compact area. The direct measurements of avalanche morphology reported below in Sec.~\ref{sec:spatial} and in the Supplementary Material~\cite{SI} show that the accompanying displacement field is itself a compact fractal, a result we reconcile with the length-controlled cutoff there.

%%%%%%%%%%%%%%%%%%%%%%%%%%%%%%%%%%%%%%%%%%%%%%%%%%%%%%%%%%%%
\subsection{Avalanche Duration Distributions}
%%%%%%%%%%%%%%%%%%%%%%%%%%%%%%%%%%%%%%%%%%%%%%%%%%%%%%%%%%%%

\begin{figure}[t!]
    \centering
    \includegraphics[width=\columnwidth]{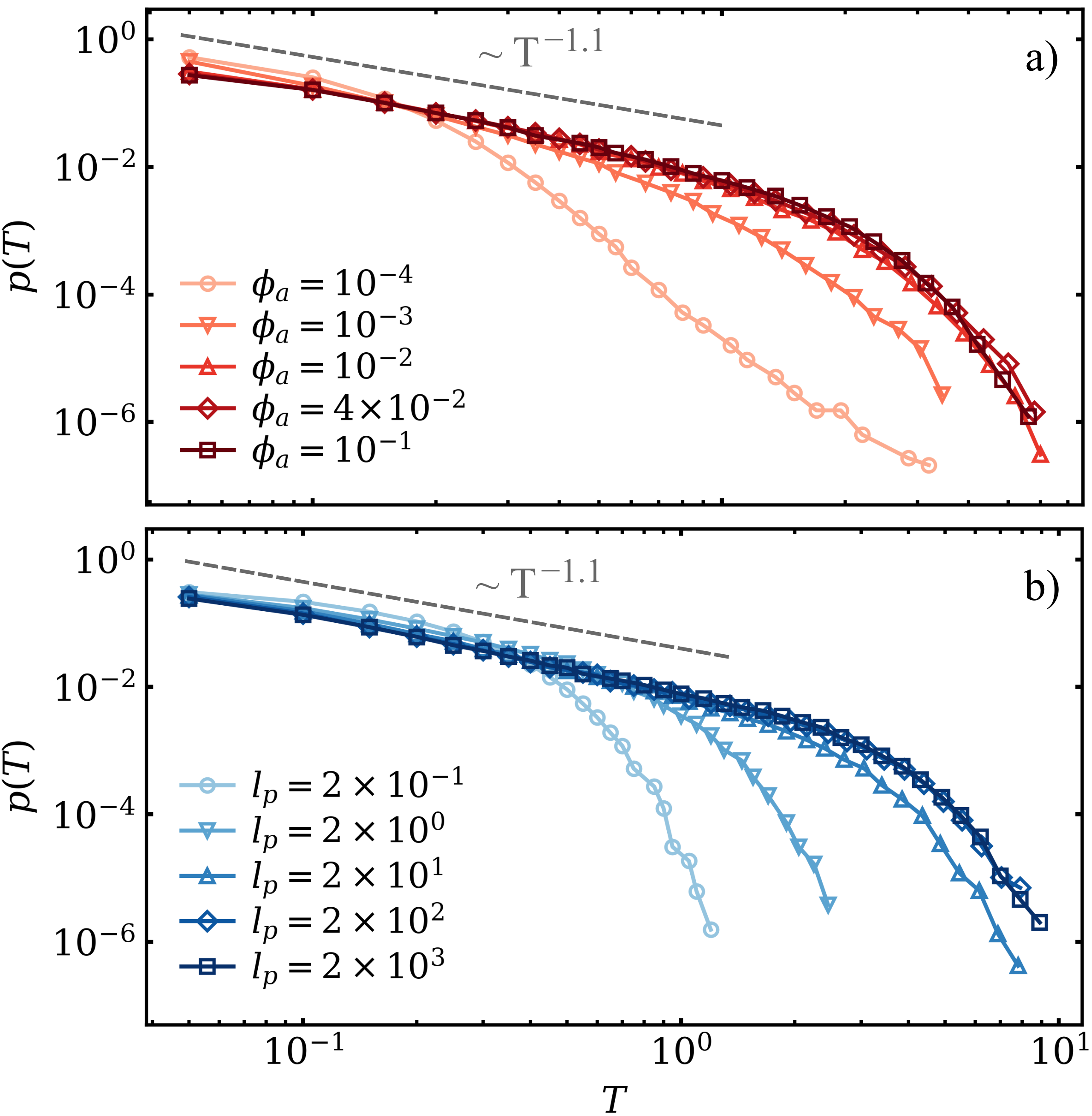}
    \caption{Avalanche duration distributions $p(T)$ upon varying $\phi_a$ at fixed $l_p = 10$ (top panel), and upon varying $l_p$ at fixed $\phi_a = 10^{-1}$ (bottom panel). Increasing either $\phi_a$ or $l_p$ drives the distributions from an exponential decay toward a power-law $p(T) \propto T^{-\tau_{_T}}$ with $\tau_{_T} \simeq 1.1$ (dashed gray lines). For sufficiently large values ($\phi_a \ge 10^{-2}$ or $l_p \ge 2 \times 10^2$) the distributions overlap, sharing an identical cutoff timescale.}
    \label{fig:avalanche_duration_distributions}
\end{figure}

The duration distributions $p(T)$ (Fig.~\ref{fig:avalanche_duration_distributions}) follow a similar crossover. Increasing either $\phi_a$ or $l_p$ drives the distributions from a fast decay toward a power-law $p(T)\propto T^{-\tau_{_T}}$ with $\tau_{_T} \simeq 1.1$, terminated by a cutoff at large durations.

At the lowest active fraction $\phi_a = 10^{-4}$ [Fig.~\ref{fig:avalanche_duration_distributions}(a)], the distribution decays rapidly, confirming that highly dilute populations sustain only short-lived avalanches. A distinction from $p(S)$ then emerges. Whereas the size cutoffs shift systematically with $\phi_a$, the $p(T)$ curves overlap and share a common duration cutoff for $\phi_a \gtrsim 10^{-2}$.
Varying $l_p$ at fixed $\phi_a$ [Fig.~\ref{fig:avalanche_duration_distributions}(b)] shows that, unlike $p(S)$, whose cutoff was nearly insensitive to intermediate values of $l_p$, the duration distributions remain sensitive to persistence and collapse only for $l_p \gtrsim 2\times10^{2}$, coinciding with the boundary of the void-forming regime.

\begin{figure}[t!]
    \centering
    \includegraphics[width=\columnwidth]{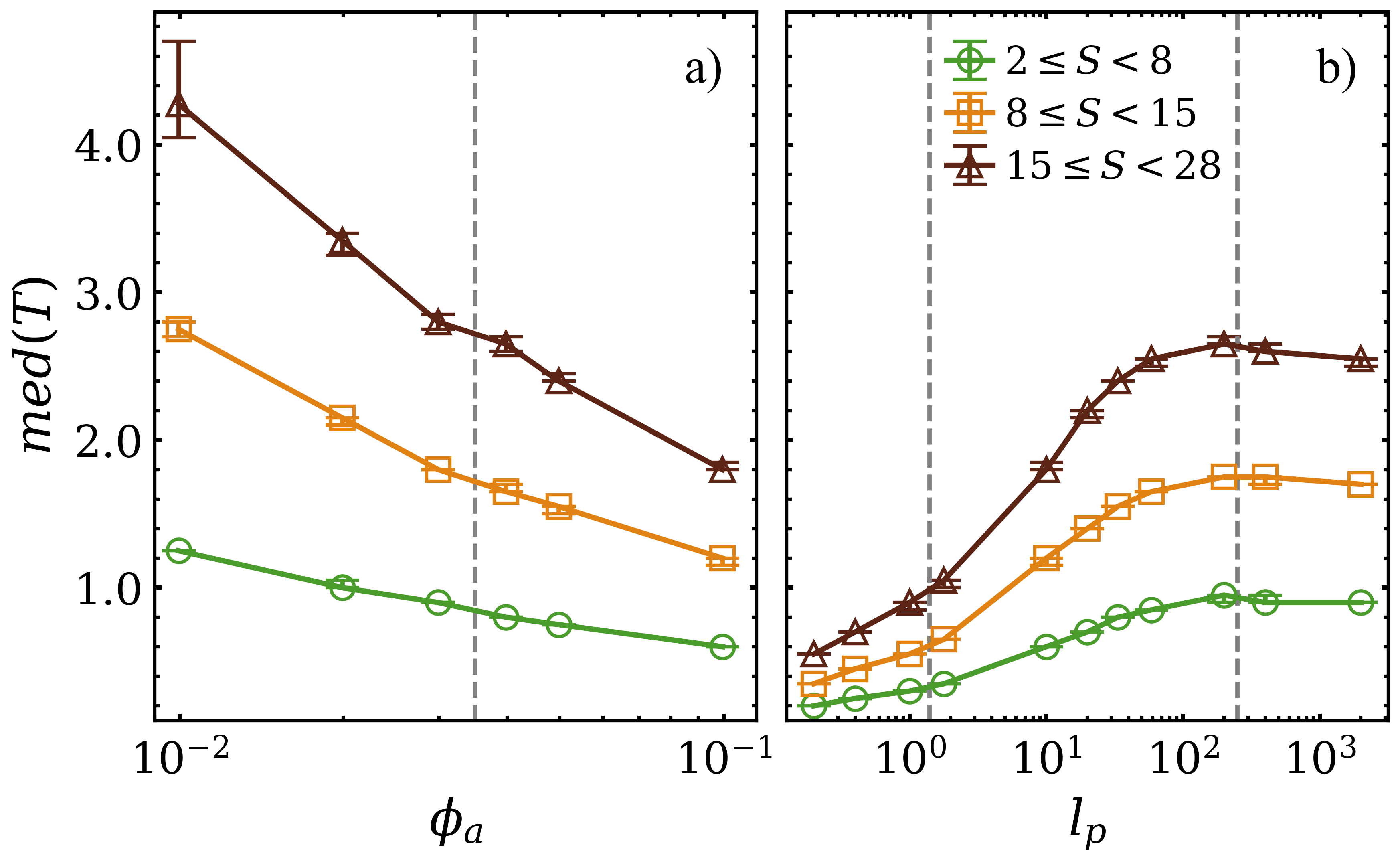}
    \caption{Median avalanche duration $\text{med}(T)$ conditioned on three fixed avalanche-size windows (see legend). (a) $\text{med}(T)$ versus $\phi_a$ at fixed $l_p = 10$: for a given size, increasing $\phi_a$ shortens the duration. Curves start at $\phi_a \ge 10^{-2}$, since smaller $\phi_a$ yield no events in the analyzed windows. (b) $\text{med}(T)$ versus $l_p$ at fixed $\phi_a = 10^{-1}$: for a given size, the duration increases with $l_p$ before saturating, an effect that strengthens for larger avalanches. Error bars denote $95\%$ block-bootstrap confidence intervals ($10^3$ replicates, blocks of $10^3$ time units). Vertical dashed lines mark the phase boundaries of Ref.~\cite{Janzen2026}.}
    \label{fig:median_duration}
\end{figure}

To resolve the temporal dynamics, we analyze the median duration conditioned on three fixed size windows, chosen in the tails where large-scale events occur, as a function of $\phi_a$ [Fig.~\ref{fig:median_duration}(a)] and $l_p$ [Fig.~\ref{fig:median_duration}(b)]. 
Increasing $\phi_a$ at fixed $l_p$ systematically shortens the median duration across all windows (the analysis is restricted to $\phi_a\gtrsim10^{-2}$, where such events are statistically accessible). A higher density of dopants reduces the average distance between triggering sites, allowing the same stress to be released more cooperatively and hence faster. This acceleration explains the collapse of $p(T)$ at high $\phi_a$. Increasing $\phi_a$ shifts the $p(S)$ cutoff to larger sizes [Fig.~\ref{fig:avalanche_size_distributions}(a)] while making avalanches of a given size faster [Fig.~\ref{fig:median_duration}(a)], and for $\phi_a\gtrsim10^{-2}$ the two trends compensate so that $p(T)$ becomes essentially independent of $\phi_a$.

Conversely, the median duration increases with $l_p$ before flattening into a plateau [Fig.~\ref{fig:median_duration}(b)]. Since $p(S)$ is nearly unaffected by $l_p$ [Fig.~\ref{fig:avalanche_size_distributions}(b)], the evolution of $p(T)$ here is controlled by the duration of avalanches of a given size, which last longer under more persistent driving. At small $l_p$ a dopant reorients many times within a single avalanche (the number of reorientations during an event of duration $T$ is $T/\tau_r$, and $\tau_r = l_p/v_0$ is then shorter than the measured durations), so the rapidly reorienting dopants act as an effective high-frequency noise that continuously perturbs the matrix and accelerates the succession of plastic events within a cascade. As $l_p$ grows, this effective noise weakens and the triggering of successive events slows. Beyond $l_p\sim2\times10^2$, which marks the transition to the void-formation regime~\cite{Janzen2026}, the forcing is effectively frozen over the time scales on which the host rearranges, the cascade decouples from the active fluctuations, and the median duration saturates at a pace set by the host alone. 
The persistence time at this crossover, $\tau_r=l_p/v_0\sim2\times10^2$,
is far longer than the elastic time $(\mu k)^{-1}$ and than the duration of individual avalanches, so the relevant host time scale is the slow structural reorganization that also controls void nucleation~\cite{Janzen2026}, rather than the elastic response.

These two trends show that the size and the duration of an avalanche are controlled independently, and map onto the distinction between ``static'' and ``dynamical'' avalanche exponents drawn for elastoplastic models by Ferrero and Jagla~\cite{FerreroJagla2019}. The size exponent, robust across protocols, is governed by the host, whereas the internal temporal development is sensitive to the drive, sped up by $\phi_a$ and slowed by $l_p$, much as finite-rate driving reshapes avalanche durations and temporal profiles at fixed host mechanics~\cite{Liu2016,SalernoRobbins2013}. The complementary size--duration relation $\langle S\rangle\propto T^\gamma$, with $\gamma\simeq1.6$ across all parameters, is reported in the Supplementary Material~\cite{SI}.

%%%%%%%%%%%%%%%%%%%%%%%%%%%%%%%%%%%%%%%%%%%%%%%%%%%%%%%%%%%%
\subsection{Active and Passive Contributions to Avalanches}
%%%%%%%%%%%%%%%%%%%%%%%%%%%%%%%%%%%%%%%%%%%%%%%%%%%%%%%%%%%%

Decomposing the stress into the species-resolved contributions $\sigma^A$ and $\sigma^P$ lets us isolate the role of each population in a way that all-active studies cannot.

\begin{figure}[t!]
    \centering
    \includegraphics[width=\columnwidth]{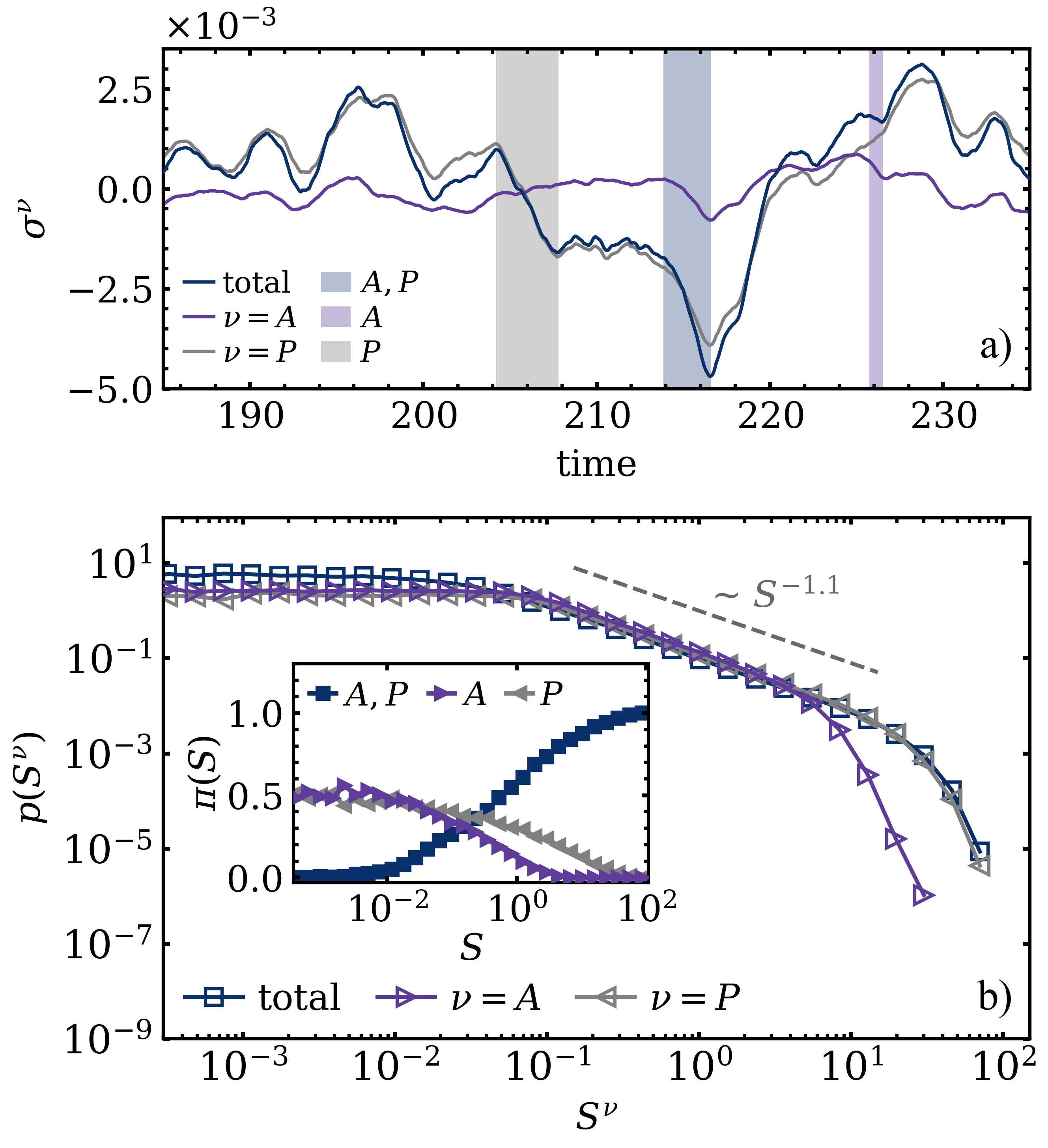}
    \caption{Active and passive contributions to the avalanches at $\phi_a = 10^{-1}$ and $l_p = 2\times10^{3}$.
    (a) Shear stress $\sigma^{\nu}$ as a function of time $t$: total stress (dark blue) and contributions of the active ($\nu = A$, purple) and passive ($\nu = P$, grey) particles. The shaded bands mark one example of each type of avalanche: both species unload (both $A$ and $P$, blue); only the active particles unload while the passive ones are loaded ($A$, purple); only the passive particles unload while the active ones are loaded ($P$, grey).
    (b) Distribution $p(S^{\nu})$ of the avalanche size $S^{\nu}$, i.e., the stress drop carried by species $\nu$ during an avalanche, for the active (purple open right-triangles) and passive (grey open left-triangles) particles, compared with the distribution of the total size $S$ (dark blue open squares, same curve as in Fig.~\ref{fig:avalanche_size_distributions}). The dashed line is a guide to the eye $\sim S^{-1.1}$.
    Inset: fraction $\Pi(S)$ of avalanches of total size $S$ belonging to each type.
    }
    \label{fig:active_passive}
\end{figure}

We compute the partial avalanche statistics $p(S^\nu)$ for the active ($\nu=A$) and passive ($\nu=P$) sets, using the partial stresses of Eq.~\eqref{eq:sigma_nu} over the avalanche windows detected on the total stress (Sec.~\ref{sec:stress_signal}). Figure~\ref{fig:active_passive}(a) shows the partial stresses $\sigma^A$ and $\sigma^P$ together with the total one on a portion of the stress signal.

Figure~\ref{fig:active_passive}(b) shows the partial avalanche size distributions $p(S^A)$ and $p(S^P)$ compared to the total distribution $p(S)$ for $\phi_a = 10^{-1}$ and $l_p = 2\times10^3$, the latter being the same curve shown in Fig.~\ref{fig:avalanche_size_distributions}. We show this state because the separation between the active and passive cutoffs is clearest at large persistence; the partial distributions across the full $\phi_a$ and $l_p$ scans are reported in the Supplementary Material~\cite{SI}.
We observe that both partial distributions follow the trend of the global one, preserving a power-law region with exponent $\tau_S \simeq 1.1$, yet the distribution $p(S^A)$ of the active contribution $S^A$ is cut off earlier. 
This is to be expected in the range of $\phi_a$ we work with, since active particles are spread throughout the system and alone they cannot build correlations beyond a certain typical `cluster' size. 
Passive particles, instead, are much more numerous and at these densities they essentially form a single cluster that spans the system, which is why the cutoff of $p(S^P)$ basically coincides with that of $p(S)$. 
In the correlated regime, the cutoffs of all three distributions shift together as $\phi_a$ and $l_p$ are varied, while the gap between the active and passive ones remains roughly constant (see Supplementary Material~\cite{SI}).

We can further disentangle the way in which active and passive particles contribute to the avalanche statistics by noticing that, since the restricted $\Delta\sigma^\nu$ can be negative, each avalanche falls into one of three classes: both species unload ($\Delta\sigma^A > 0$, $\Delta\sigma^P > 0$), only the active particles unload ($\Delta\sigma^A > 0$, $\Delta\sigma^P \le 0$), or only the passive particles unload ($\Delta\sigma^A \le 0$, $\Delta\sigma^P > 0$); an example of each class is marked by the shaded bands in Fig.~\ref{fig:active_passive}(a).
For the system explored in Fig.~\ref{fig:active_passive}, only $\sim 46\%$ of the avalanches involve a simultaneous release of stress in both species. In the remaining ones, one species unloads while the other is loaded: the active particles alone unload in $\sim 22\%$ of the events, and the passive particles alone in the remaining $\sim 32\%$. These fractions become $\sim 61\%$, $\sim 13\%$ and $\sim 26\%$, respectively, when only avalanches with $S>10^{-1}$ are considered, that is, avalanches in the scaling regime, beyond the noise encoded in the initial plateau.
The inset of Fig.~\ref{fig:active_passive}(b) shows how these fractions evolve with the total avalanche size $S$.
Avalanches led by stress releases of the active particles, while the passive host increases its stress, are typically very small. 
On the other hand, avalanches originating from passive particles that release more stress than the active dopants absorb can be somewhat larger, but their frequency also vanishes with increasing $S$. Only when both species release stress simultaneously do we reach the largest avalanche sizes $S$ observed for a given system size; these avalanches also dominate most of the scaling regime ($S>10^{-1}$).
The same picture holds across all state points (Supplementary Material~\cite{SI}): the fraction of avalanches in which both species unload grows with $S$ and dominates the largest avalanches, whereas the fraction of active-only avalanches increases with $\phi_a$ and depends only weakly on $l_p$.

%%%%%%%%%%%%%%%%%%%%%%%%%%%%%%%%%%%%%%%%%%%%%%%%%%%%%%%%%%%%
\subsection{Spatial Structure of Avalanches}
\label{sec:spatial}
%%%%%%%%%%%%%%%%%%%%%%%%%%%%%%%%%%%%%%%%%%%%%%%%%%%%%%%%%%%%
\begin{figure*}
    \centering
    \includegraphics[width=\textwidth]{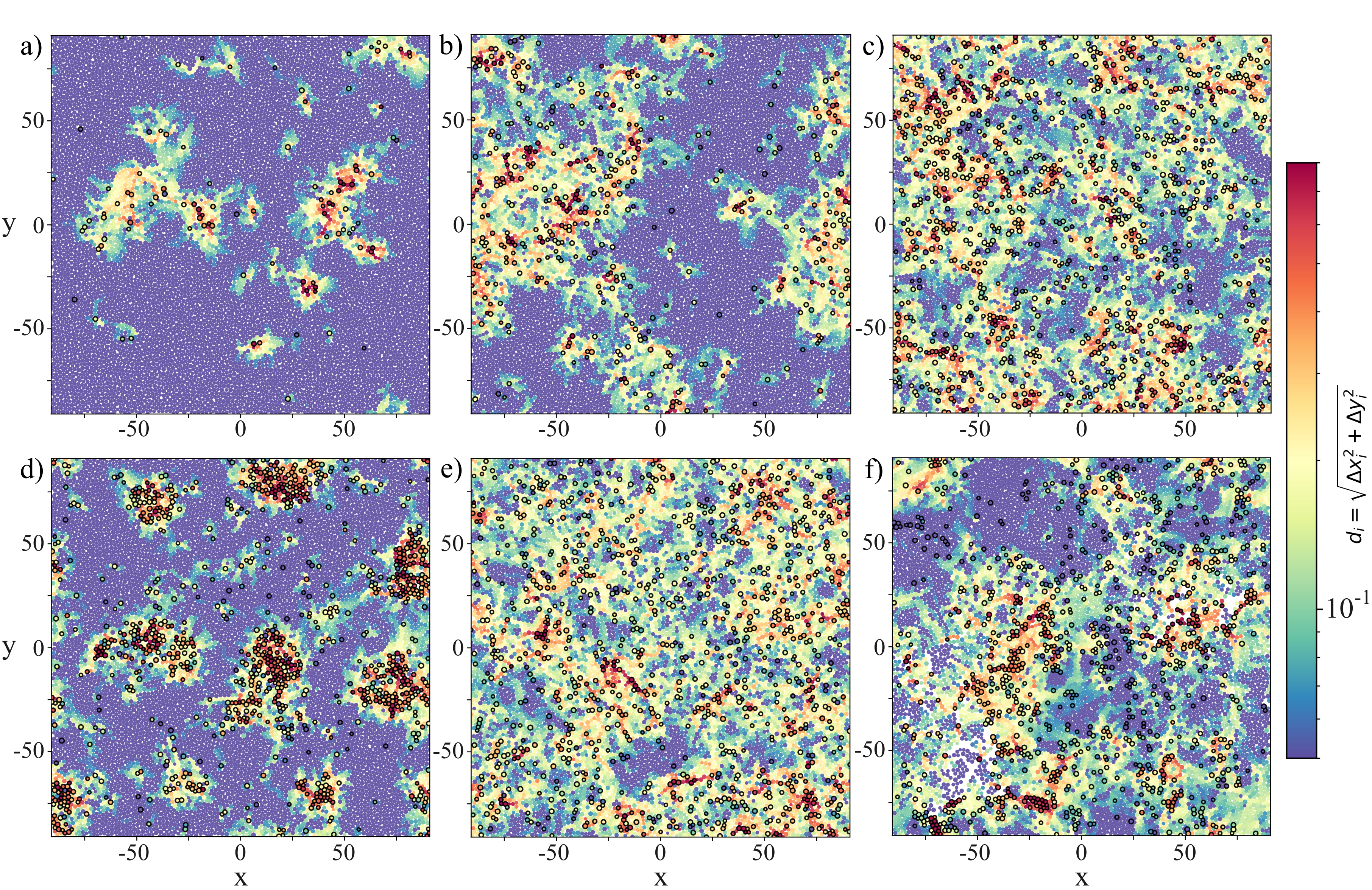}
    \caption{Representative snapshots at the end of avalanches of comparable size ($S \approx 25$). Top: varying $\phi_a$ at fixed $l_p = 10$: (a) $\phi_a = 10^{-2}$, (b) $4 \times 10^{-2}$, (c) $10^{-1}$. Bottom: varying $l_p$ at fixed $\phi_a = 10^{-1}$: (d) $l_p = 2 \times 10^0$, (e) $2 \times 10^1$, (f) $2 \times 10^3$. Each particle is colored by its accumulated displacement $d_i$ over the avalanche (color bar saturated at $d_i = 0.8$). Black outlines mark active particles.}
    \label{fig:snapshots_active_dopants}
\end{figure*}

To resolve the spatial structure of the stress releases, for each parameter set we select a representative avalanche of comparable size ($S\approx25$) and color each particle by its accumulated displacement $d_i$ (Fig.~\ref{fig:snapshots_active_dopants}, with remaining cases in the Supplementary Material~\cite{SI}). The displacement, rather than a thresholded mobility field, provides a threshold-free readout of the rearrangement, complementary to the non-affine measure $D^2_{\min}$ of Falk and Langer~\cite{FalkLanger1998}.

The top panels show a structural transition with $\phi_a$ at fixed $l_p=10$. At low concentration [Fig.~\ref{fig:snapshots_active_dopants}(a)] the morphology consists of localized, disconnected mobile clusters (bright regions) embedded in a quiescent matrix (purple), concentrated around the active dopants (black circles), which act as localized plastic triggers. As $\phi_a$ increases [Fig.~\ref{fig:snapshots_active_dopants}(b)] these domains grow, and at the highest concentration [Fig.~\ref{fig:snapshots_active_dopants}(c)] they merge into a single, system-spanning cooperative region. The bottom panels show a non-monotonic change with $l_p$ at fixed $\phi_a=10^{-1}$. At low $l_p$ the clusters are compact and tightly bound to the dopants [Fig.~\ref{fig:snapshots_active_dopants}(d)]. At intermediate $l_p$ the cascades reach their largest spatial extent and span the system [Fig.~\ref{fig:snapshots_active_dopants}(e)]. At very high $l_p$ the structure reverts to a localized form in which the extreme persistence opens structural voids~\cite{Janzen2026} and plastic activity localizes around these cavities [Fig.~\ref{fig:snapshots_active_dopants}(f)]. The mobile regions organize into extended, correlated structures rather than independent single-particle moves, a morphology that resembles the long-range displacement correlations and slip-line patterns reported for sheared amorphous solids~\cite{MaloneyLemaitre2006,MaloneyRobbins2008}.

\begin{figure}[t!]
    \centering
    \includegraphics[width=\columnwidth]{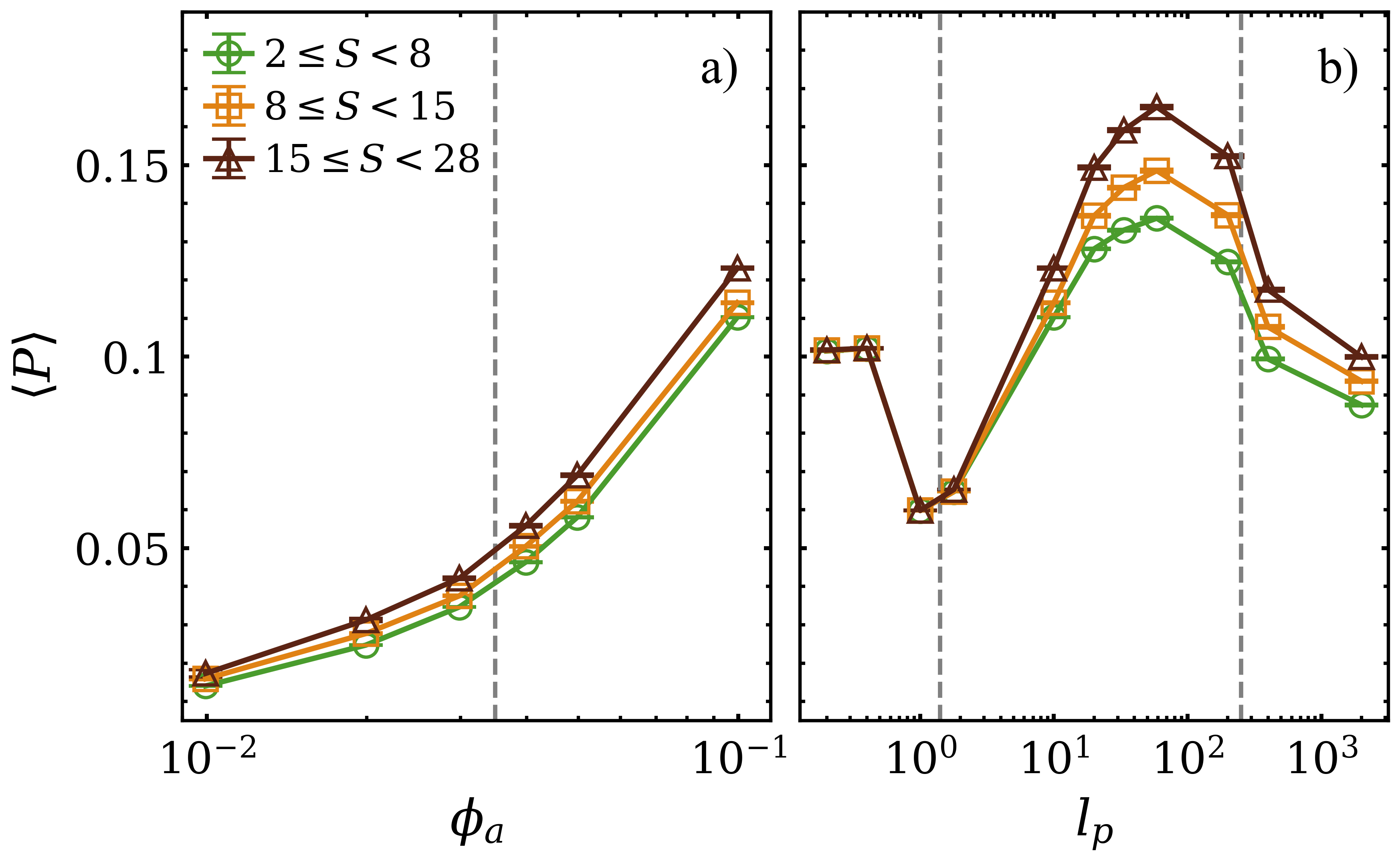}
    \caption{Spatial extent of avalanches quantified by the average participation ratio $\langle P \rangle$, computed in three avalanche-size windows (see legend; same windows as in Fig.~\ref{fig:median_duration}); error bars are standard errors. (a) $\langle P \rangle$ versus $\phi_a$ at fixed $l_p = 10$: the spatial extent grows monotonically with $\phi_a$, reflecting the crossover from isolated to more extended mobile clusters. (b) $\langle P \rangle$ versus $l_p$ at fixed $\phi_a = 10^{-1}$: for $l_p \lesssim 2$ $\langle P \rangle$ is independent of $S$, it then increases within the fluidized window, peaking at intermediate persistence ($l_p \approx 10^1$--$10^2$), and decreases again at large $l_p$ due to local segregation. Vertical dashed lines mark the phase boundaries of Ref.~\cite{Janzen2026}.}
    \label{fig:participation}
\end{figure}

To quantify these trends we compute the average participation ratio $\langle P\rangle$ of avalanches in three size windows (Fig.~\ref{fig:participation}). Varying $\phi_a$ at fixed $l_p=10$ [Fig.~\ref{fig:participation}(a); restricted to $\phi_a\ge10^{-2}$, where all three windows are populated], $\langle P\rangle$ grows monotonically for all sizes, from $\approx0.015$ at $\phi_a=10^{-2}$, consistent with isolated, localized clusters, to $\approx0.12$ at $\phi_a=10^{-1}$, as the growing density of dopants lets their local perturbations overlap and span larger portions of the system. Varying $l_p$ at fixed $\phi_a=10^{-1}$ [Fig.~\ref{fig:participation}(b)], the response is non-monotonic. For $l_p\lesssim2$, $\langle P\rangle$ takes the same value in all size windows: the participation is then set by the background displacement field rather than by the avalanche itself, as expected when rapidly reorienting active forces act as an effective thermal bath~\cite{mandal2016active,berthier2019glassy}. Its minimum, at $l_p\approx1$--$2$, lies at the lower phase boundary of Ref.~\cite{Janzen2026}. Within the fluidized window, $\langle P\rangle$ increases and becomes size dependent, larger avalanches involving a larger fraction of the system, and reaches a maximum at $l_p\approx50$, where the coherent driving time is comparable to the time over which the host reorganizes structurally. Beyond the upper boundary, extreme persistence produces local structural segregation and void nucleation~\cite{Janzen2026}, which localizes plastic activity around the cavities and lowers $\langle P\rangle$ again. The non-monotonic spatial cooperation thus tracks the same finite-persistence yielding window identified for active particles more broadly~\cite{Goswami2025}, and marks the boundary beyond which the effective-temperature picture of activity~\cite{mandal2016active,berthier2019glassy} ceases to apply.

To characterize the geometry of the rearranging regions we measure the mass--radius relation of the mobile clusters, $N_k\sim\langle R_g\rangle^{d_c}$, where $N_k$ is the number of mobile particles in a cluster and $R_g$ its radius of gyration. Across both the $\phi_a$ and $l_p$ scans we find a robust cluster dimension $d_c\simeq1.7$ (Supplementary Material~\cite{SI}), interestingly coincident with the one observed in~\cite{Roland2026}, so the mobile regions are compact fractals rather than strictly line-like objects. This value must be read together with the stress-drop cutoff $S_c\sim L^{d_f}$ with $d_f\simeq 1$, which instead points to an effectively one-dimensional scaling. The two exponents are not in conflict, because they characterize different fields. The cutoff $S_c$ measures the extensive stress \emph{released}, and its system-size scaling is set by the longest spanning length over which stress is transmitted, i.e., by the quasi-one-dimensional backbone of the cascade. In contrast, $R_g$ and $N_k$ measure the \emph{displacement footprint} that dresses this backbone, a broader elastic reorganization of fractal dimension $d_c\simeq1.7$. A line-like stress-transmission path surrounded by a more space-filling displacement cloud is precisely the morphology reported for slip lines and shear-transformation chains in sheared amorphous solids~\cite{MaloneyLemaitre2006,MaloneyRobbins2008}, and it reconciles the length-controlled cutoff with the compact-fractal participation field.

The size, duration, and spatial measurements are consistent with each other. Activity sets the initiation and internal timing of relaxation, with the dopants acting as distributed cage-breaking seeds whose density fixes the spacing between nucleation sites and whose persistence tunes the pace of the cascade, while the amorphous host, through anisotropic long-range stress redistribution, governs how each release propagates and hence the exponent of the scale-free size statistics. This separation is also consistent with the observation that active and sheared systems share their avalanche exponents while their steady-state rheology differs~\cite{Villarroel2021,WieseKroyLevis2023,RelmucaoLeiva2025}.

%%%%%%%%%%%%%%%%%%%%%%%%%%%%%%%%%%%%%%%%%%%%%%%%%%%%%%%%%%%%
\section{Conclusions}
%%%%%%%%%%%%%%%%%%%%%%%%%%%%%%%%%%%%%%%%%%%%%%%%%%%%%%%%%%%%

We studied the mechanical relaxation of a dense amorphous host doped with a dilute set of persistent active particles, across the regimes of dynamic arrest, homogeneous fluidization, and void formation. The apparently homogeneous fluidized state does not relax as a simple activated liquid. It relaxes through intermittent, avalanche-like stress-release events whose size and duration distributions cross over from exponential to power-law, $p(S)\propto S^{-\tau_S}$ with $\tau_S\simeq1.1$ and $p(T)\propto T^{-\tau_{_T}}$ with $\tau_{_T}\simeq 1.1$, as either the active fraction or the persistence length increases. Thermal dopants lacking persistence produce purely exponential distributions throughout, establishing that scale-free avalanche dynamics requires coherent stress accumulation driven by persistent forcing.

The central result is a separation of roles between the drive and the host. The active fraction sets the spatial scale of relaxation. Given a fix system size, the avalanche-size cutoff scales as $S_c\sim N_a^{1/2}\sim L/\ell_a$, so the largest events are bounded by the number of dopant spacings across the system rather than by the persistence of the forcing, and the participation ratio grows with $\phi_a$ as isolated mobile clusters merge into system-spanning ones. Persistence instead controls the internal timing of relaxation. It leaves the size statistics nearly unchanged while lengthening avalanches of a given size, until the persistence time exceeds the structural reorganization time of the host near the void-formation boundary of Ref.~\cite{Janzen2026} and the dynamics decouples from the active fluctuations. Because the size exponent is set by the host and the timing by the drive, avalanche exponents measured under active and shear driving can agree~\cite{Morse2021,Villarroel2024,Roland2026} while the dynamical exponents vary ~\cite{FerreroJagla2019}.

We interpret the active dopants as spatially distributed cage-breaking seeds and the amorphous host as the medium that propagates the resulting stress release. 
Once an avalanche of plastic events is triggered by one of those cage-breaking seeds, the entire system responds as a dense amorphous solid with plastic events mediated by an Eshelby-like propagator happening elsewhere.
As a matter of fact, we observe a finite-size scaling for the avalanche size distributions consistent with slip-line avalanches and shear-transformation-chain organization of sheared amorphous solids~\cite{MaloneyLemaitre2006,MaloneyRobbins2008,Nicolas2018}.
Yet, the amount of injected strain, controlled by the fraction of active particles impacts in the size of the largest avalanches observed at a fix system-size, yielding a square-root scaling with $N_a$ of the size cutoff, interestingly contrasting with the case of increasing strain-rate in purely passive amorphous systems, that leave the size cutoff essentially unchanged~\cite{Liu2016}.
Two consequences can be tested directly. Varying the active number $N_a$ at fixed packing fraction should shift the size cutoff as $N_a^{1/2}$, and the spatial cooperation should peak in the finite-persistence window bracketed by the fluidization and void-formation boundaries. Both are accessible in active colloidal dopants embedded in dense passive packings and in motile bacteria dispersed in crowded media, where the dopant concentration and persistence can be tuned independently~\cite{Goswami2025,Shee2025dopants}.

Our estimate $\tau_S\simeq1.1$ carries the protocol and damping dependence documented for overdamped particle models~\cite{Salerno2012,SalernoRobbins2013,Liu2016}, and should not be read as identifying a distinct universality class. The square-root cutoff scaling is evidence for lower-dimensional propagation but not proof of strictly one-dimensional avalanches. Beyond the mass--radius scaling of the mobile clusters reported here, measurements of the anisotropy of the stress-release field are required to distinguish line-like elastic cascades from other non-compact mechanisms. 
Connecting the avalanche statistics reported here to the steady-state rheology of active and active--passive systems~\cite{Villarroel2021,WieseKroyLevis2023,RelmucaoLeiva2025} would clarify how the stress released in avalanches translates into flow.

\section{Acknowledgements}
CL and EEF acknowledge the support from the Simons-Balseiro Exchange Program in the framework of the Simons Foundation Targeted Grant to Institutions SFI-MPS-T-Institutes-00011985. CV acknowledges fundings  PID2025-170781NB-C21 and PID2022-140407NB-C21 funded by MCIN/AEI /10.13039/501100011033 and FEDER, UE. DMF  acknowledges support from MINECO (MCIN/AEI Grant No. PID2023-148991NA-I00) and the Ramon y Cajal Program of the Spanish Ministry of Science and Innovation (Grant No. RYC2023-043002-I).

%%%%%%%%%%%%%%%%%%%%%%%%%%%%%%%%%%%%%%%%%%%%%%%%%%%%%%%%%%%%
\bibliography{landi_et_al_rev1}
%\bibliography{references}
%%%%%%%%%%%%%%%%%%%%%%%%%%%%%%%%%%%%%%%%%%%%%%%%%%%%%%%%%%%%

\end{document}